 \documentclass[twocolumn,pra,superscriptaddress,showpacs,floatfix]{revtex4}
\usepackage{graphicx}
\usepackage[latin1]{inputenc}
\usepackage[]{amsmath}
\usepackage{dcolumn}
\usepackage{bm}
\usepackage{tabularx}
\usepackage{color}

\graphicspath{.}

\begin{document}
\title{Multiconfiguration Dirac-Hartree-Fock calculations of
atomic  electric dipole moments
 of $^{225}$Ra, $^{199}$Hg, and $^{171}$Yb}
\author{Laima Rad\v zi\= ut\.e}
\author{Gediminas Gaigalas}
\affiliation{Vilnius University,
             Institute of Theoretical Physics and Astronomy, \\
             A.~Go\v{s}tauto 12, LT-01108, Vilnius, Lithuania}
\author{Per J\"onsson}
\affiliation{Group for Materials Science and Applied Mathematics,\\
             Malm\"o University, S-20506, Malm\"o, Sweden}
\author{Jacek Biero\'n}
\affiliation{Instytut Fizyki imienia~Mariana Smoluchowskiego,
             Uniwersytet Jagiello{\'n}ski,
             Krak{\'o}w, Poland}
%
\date{\today}
\begin{abstract}
The multiconfiguration Dirac-Hartree-Fock (MCDHF) method has been employed
to calculate atomic electric dipole moments (EDM) of
$^{225}$Ra, $^{199}$Hg, and $^{171}$Yb.
For the calculations of the matrix elements we extended
the relativistic atomic structure package GRASP2K~\cite{grasp2K:2013}. 
The extension includes programs to evaluate 
matrix elements of $(P,T)$-odd e-N
tensor-pseudotensor and pseudoscalar-scalar interactions,
the atomic electric dipole interaction,
the nuclear Schiff moment, and the interaction of the electron electric dipole 
moment with nuclear magnetic moments.
The interelectronic interactions were accounted for
through valence and core-valence electron correlation effects.
The electron shell relaxation was included 
with separately optimised wave functions of opposite parities.
\end{abstract}

%
\pacs{ 11.30.Er, 32.10.Dk, 31.15.A-, 24.80.+y}


\maketitle

\section{Introduction}
\label{sectionIntroduction}

The existence of a non-zero permanent electric dipole moment
(EDM) of an elementary particle or a composite system of
particles would violate
time reversal symmetry (T), as well as
the combined charge conjugation and parity symmetry (CP),
due to the CPT theorem~\cite{KhriplovichLamoreaux}.
One of the principal motivations behind
the experimental searches of EDMs 
is to shed light on the observed 
matter-antimatter asymmetry in the Universe, which in turn
is linked to an asymmetry in the Big Bang baryon-antibaryon
production.
The standard model (SM) of elementary particles cannot explain
the matter-antimatter asymmetry in the Universe, 
as SM predicts sources of CP violation
(and of EDMs) several orders of magnitude weaker
than those needed to account for the observed baryon numbers.
This leads to proliferation of the extensions to the 
standard model.
Some of these extensions predict larger EDMs, sometimes within 
the reach of current experiments.
The experimental searches have not yet detected a non-zero EDM,
but they continue to improve the limits on EDMs of individual
elementary particles, as well as limits on CP-violating
interactions, usually parametrized by the 
interaction constants 
$C_{T}$, and
$C_{P}$
(see section~\ref{sectionEDMtheory} for details
and the Table~II in the reference~\cite{Griffith:2009} for a summary).
These limits 
constrain the theoretical extensions of the standard model of elementary
particles.
In recent years these constraints have been set by the measurements
of EDMs of 
neutrons~\cite{Baker:2006},
electrons in a paramagnetic atom
(a thallium atom experiment~\cite{Regan:2002}),
electrons in a  diamagnetic atom
(mercury atom~\cite{Griffith:2009}),
and in TlF and  YbF molecules~\cite{Hinds:2011}.
The search for EDMs is not restricted to the above
species, though --- see
e.g.~\cite{RobertsMarciano:2009}.

The search for a permanent electric dipole moment
of an elementary particle, or a composite system of
particles
(see~\cite{KhriplovichLamoreaux}, or a recent
reference~\cite{DzubaFlambaum:2012} for a review),
is a challenge, not only for experiments, but also for theories
of composite systems.
Heavy atoms are excellent examples
of composite systems with large EDMs, due to the existence
of mechanisms which may induce atomic EDMs several orders
of magnitude larger than an intrinsic particle EDM.
%
In the present paper we computed the EDMs of three diamagnetic
atoms, 
$^{225}$Ra, $^{199}$Hg, and $^{171}$Yb.
The purpose of the present paper is fourfold.
Firstly, we tested the newly developed 
programs to evaluate 
matrix elements of  $(P,T)$-odd e-N
tensor-pseudotensor and pseudoscalar-scalar interactions,
the atomic electric dipole interaction,
the nuclear Schiff moment, and the interaction of the electron electric dipole 
moment with nuclear magnetic moments.
Secondly, 
we generated the atomic wave functions in several different approaches,
in order to test the dependence of the calculated atomic EDMs
on options available in the GRASP2K~\cite{grasp2K:2013} implementation
of the MCDHF method. 
The approaches depended on the choice of variational energy functional
(Average Level versus Optimal Level,
 with different numbers of optimised levels),
the choice of wave functions built on a common orbital set or several separately optimised orbital sets,
in the latter case biorthogonal transformations of wave functions
 had to be applied,
as well as on specific methods of one-electron orbital generation.
All these approaches are discussed in more detail in
section~\ref{sectionEnergyFunctionals},~\ref{sectionVirtualOrbitals},~\ref{sectionNonOrtho} and 
\ref{sectionExtendedOptimalLevels}
and presented in Tables 
\ref{ortho-vs-nonortho},
\ref{EOL_TPT_Ra_Ex}, and
\ref{tableRaTPTenergies}.
Thirdly,
we sequentially generated several layers of virtual 
(correlation) orbitals
for each of the three elements
and observed the effects of electron correlation on atomic EDMs.
All valence and core-valence electron correlation effects were
included through single and restricted double electron substitutions
from core-valence to virtual orbitals.
And finally, we provide independently calculated
atomic EDMs in 
$^{225}$Ra, $^{199}$Hg, and $^{171}$Yb,
and compare our results with those of other authors.
Our results,
presented in the
Tables~\ref{tableRaHgYbTPTresults}, 
\ref{tableSPS},
\ref{tableSchiff}, and
\ref{tableeEDM_M},
were obtained within the  
multiconfiguration Dirac-Hartree-Fock (MCDHF) method,
using the relativistic atomic structure package GRASP2K~\cite{grasp2K:2013},
which, to the best of our knowledge,
and with the exception of one paper~\cite{Bieron:RaSchiff:2009}
on the Schiff moment in radium,
has been employed for the first time
in the calculations of
matrix elements of $(P,T)$-odd e-N
tensor-pseudotensor and pseudoscalar-scalar interactions,
nuclear Schiff moment, and interaction of electron electric dipole 
moment with nuclear magnetic moments.

%
The three atoms
$^{225}$Ra, $^{199}$Hg, and $^{171}$Yb,
have been chosen on the grounds that they have similar
valence shell structure.
All these elements are diamagnetic, with closed outer $s$ shell
($^{225}$Ra $ 6p^6 7s^2 $,
$^{199}$Hg $ 5d^{10} 6s^2 $, and
$^{171}$Yb $ 4f^{14} 6s^2 $).
In the future we will be able extend these calculations to
closed-\textit{p}-valence-shell atoms, as well as
to any other, closed- or open-shell system.
Our current MCDHF machinery~\cite{grasp2K:2013} is robust enough to deal with
electron correlation effects in
arbitrary atomic systems, including the lanthanides and actinides.

%
\section{EDM theory}
\label{sectionEDMtheory}

The interactions which
mix atomic states of different parities
and
induce a static electric dipole moment of an atom
are quite weak. 
Therefore an atomic wave function can be expressed as
\begin{eqnarray}
\label{eq:mixed_parity}
   \widetilde{\Psi}  \left( J M_J \right) \; & = &
   a \Psi  \left( \gamma P J  M_J \right) \; + \;
 \nonumber \\ [1ex]
  &  &  \hspace*{-0.2cm}
   \sum_{i} ^{} \; b_i \; 
   \Psi \left( \gamma _i  (-P) J_i  M_{J_{i}} \right) \, 
\end{eqnarray}
where the coefficient $a$ of the dominant contribution can be set to 1.
The expansion coefficients of opposite parity ($-P$) admixtures,
$b_{i}$, 
can be found using first order perturbation theory:
\begin{eqnarray}
\label{eq:b_factor}
   b_i \; = \;
   \frac{
   \left< \Psi \left( \gamma_i  (-P) J_i  M_{J_{i}} \right) | \hat{H}_{int}
    |\Psi  \left( \gamma  P J M_J \right) \right>}
   {E \left( \gamma P J \right) \; - \; E \left( \gamma_i (-P) J_i \right)} 
   \, .
\end{eqnarray}
$\hat{H}_{int}$ represents the Hamiltonian of the $(P,T)$-odd interaction,
which mixes states of opposite parities.
The mixed-parity state of a particular atomic level $^{2S+1}L_{J}$
induces a static EDM of an atom:
\begin{eqnarray}
\label{eq:DA}
   d_{at}^{int}  = 
   \left< \widetilde{\Psi}  \left( \gamma J M_{J} \right) | \hat{D}_{z}
   | \widetilde{\Psi}  \left( \gamma J M_J \right) \right> 
   & = &
 \nonumber \\ [0.2cm]
   &  & \hspace*{-7cm}
   \; 2  \sum_{i}^{} b_{i}
   \left< \Psi \left( \gamma P J M_J \right) | \hat{D}_{z} |
          \Psi \left( \gamma _i (-P) J_i M_{J_{i}} \right)
   \right> ,
\end{eqnarray}
where $\hat{D}_z$ represents the $z$ projection of the electric-dipole 
moment operator. 
Eventually an atomic EDM can be written as a sum: 
\begin{eqnarray}
\label{eq:DAHint}
   d_{at}^{int}  = 
   \; 2  \sum_{i}
    \frac{ \left< 0| \hat{D}_{z} |i\right> 
   \left< i | \hat{H}_{int}    |0 \right>}
   {E_{0} \; - \; E_{i}} 
  ,
\end{eqnarray}
 where $ | 0  \rangle $ represents the ground
 state $|\Psi \left( \gamma  P J M_J \right>$,
with $J=0$ and even parity, and the summation runs over excited
states $|\Psi \left( \gamma_{i} (-P) J_{i} M_{J_{i}} \right>$,
with $J_{i}=1$ and odd parity. $E_{0}$ and $E_{i}$
are energies  of ground and excited  states, respectively.
In practice this sum needs to be truncated at some level.

Calculations of atomic EDM require evaluation of the
 matrix element of the static EDM
$\left< 0| \hat{D}_{z} |i\right>$
and the matrix element of the interactions which induced EDM in an atom
$\left< i | \hat{H}_{int}    |0 \right>$.
The operators associated with the above matrix elements are all one-particle operators.

 For the general tensor operator $\hat{T}^{k}_{q}$, the matrix element between 
states of different parity can be expressed by Wigner-Eckart theorem as:
\begin{eqnarray}
\label{eq:mat_element_T}
   \left< \Psi \left( \gamma  P J M_J \right) | \hat{T}_{0}^{k}
    | \Psi \left( \gamma_i (-P) J_i  M_{J_{i}} \right) \right> & = &
   \nonumber  \\ [1ex]
   &  & \hspace*{-6cm}
   (-1)^{J-M_{J}} \;
   \; \sqrt{2J+1} \;
   \left(
   \begin{array}{ccc}
      J    & k & J_{i} \\
      -M_J & 0 & M_{J_{i}}
   \end{array}
   \right)  
\nonumber  \\ [1ex]
&  & \hspace*{-6cm}
     \times \left[ \Psi  \left( \gamma P J \right) \| \hat{T}^k \| 
          \Psi  \left( \gamma_i (-P) J_i \right) \right]\ . 
\end{eqnarray}
Expanding the wave functions in configuration state functions (CSFs), $\Phi  \left( \gamma P J \right) $, that are built from one-electron Dirac orbitals,
see section~\ref{sectionMCDHFcalculations},
the reduced matrix elements of $\hat{T}^{k}_{q}$ can be written
\begin{eqnarray}
\label{eq:mat_element_between_ASF}
   \left[ \Psi  \left( \gamma P J \right) \| \hat{T}^{k} \| 
          \Psi  \left( \gamma_i (-P) J_i \right) \right]
   & = &
   \nonumber \\ [0.2cm]
   &  & \hspace*{-4.5cm}
 \sum_{r,s} \; c_r c_s
   \left[ \Phi  \left( \gamma_r P J \right) \| \hat{T}^{k} \| 
          \Phi \left( \gamma_s (-P) J_i \right) \right] \, 
\end{eqnarray}
where $c_{r}$ and $c_{s}$ are mixing coefficients of CSFs
 (even and odd parity, respectively).
The matrix elements between the CSFs, in turn, can be written as sums of single-particle
matrix elements
\begin{eqnarray}
\label{eq:mat_element_between_CSF}
   \left[ \Phi \left( \gamma_r P J \right) \| \hat{T}^{k} \| 
          \Phi \left( \gamma_s (-P) J_i \right) \right] & = &
   \nonumber \\ [0.2cm]
   &  & \hspace*{-4.5cm}
   \; \sum_{a,b} \; d^{k}_{ab}(rs)  
   \left[ n_a \kappa_a \| \hat{t}^{k} \| n_b \kappa_b \right].
\end{eqnarray}
In the latter expansion, the $d^{k}_{ab}(rs)$ are known as 
`spin-angular coefficients' that arise from using Racah's algebra in the 
decomposition of the many-electron matrix elements~\cite{GrantBook2007,SAI}.
The expressions (\ref{eq:mat_element_T}),
(\ref{eq:mat_element_between_ASF}) and 
(\ref{eq:mat_element_between_CSF}) 
are general and can be used for any one-particle operator.

We consider the following four mechanisms which may induce atomic EDM:
tensor-pseudotensor ($\hat{H}_{TPT}$),
pseudoscalar-scalar ($\hat{H}_{SPS}$),
Schiff moment ($\hat{H}_{SM}$), and
electron EDM interaction with nuclear magnetic field ($\hat{H}_{B}$).
The interactions, which are all of rank $k=1$, are discussed in more detail in the next sections. 
In addition the expression for the electric dipole interaction is given.

%
\subsection{The electric dipole operator}
The electric-dipole moment operator has the rank $k$=1
 in (\ref{eq:mat_element_T}), (\ref{eq:mat_element_between_ASF}), and 
(\ref{eq:mat_element_between_CSF}),
and the single-particle reduced matrix element
$\left[ n_a \kappa_a \| \hat{t}^{k} \| n_b \kappa_b \right]$ 
in equation~(\ref{eq:mat_element_between_CSF}) can be written as
\begin{eqnarray}
\label{eq:radial_integral_EDM}
   \left[ n_a \kappa_a \| \hat{d}^1 \| n_b \kappa_b \right]
    & = & 
   \nonumber \\ [0.2cm]
   &  & \hspace*{-2.9cm}
- \left[ \kappa_a \| C^{1} \| \kappa_b \right] \;
   \int_{0}^{\infty}
   \hspace*{-.3cm} \left( P_a P_b \; + \; Q_a Q_b \right) \; r \; dr,
\end{eqnarray}
where $P$ and $Q$ are large and small components of 
the relativistic radial wave functions, respectively.
The single-particle angular reduced matrix elements can be expressed as:
\begin{eqnarray}
\label{eq:d}
    \left[ \kappa_a \| C^k \| \kappa_b \right]
    & & = 
    \nonumber \\ [0.2cm]
    &  & \hspace*{-2.9cm}
    (-1)^{j_a +1/2} \; \sqrt{2j_b \; + \; 1}
       \left(
   \begin{array}{ccc}
      j_a   & k & j_b \\
      1/2 & 0 & -1/2 
   \end{array}
   \right)
      \pi \left( l_a, l_b, k\right),
\end{eqnarray}
where $\pi \left( l_a, l_b, k\right)$ is defined as:
\begin{eqnarray}
\label{eq:pi}
  \pi \left( l_a, l_b, k\right) 
& & =
  \left\{
  \begin{array}{ll}
  1 ; & \mbox{ if } l_a+k+l_b \mbox{ even,} \\ 
  0 ; & \mbox{ otherwise. }
  \end{array}
  \right.
\end{eqnarray}

%

%
\subsection{Tensor-pseudotensor interaction}
\label{sectionTPT}

One of the possible sources of the EDM in diamagnetic atoms is the
tensor-pseudotensor (TPT) interaction  between electrons and nucleons,
violating both parity (P) and time (T)-reversal invariance.
It can be expressed as

 \begin{eqnarray}
\label{eq:H_T}
   \hat{H}_{TPT}  \; = \;
   i\sqrt{2}G_{F}C_{T} \;
   \sum_{j=1} ^{N} \; \left( < \hspace*{-.1cm}\bm{ {\sigma}}_{A}
   \hspace*{-.1cm}> \cdot \, \hspace*{.01cm} \bm{\gamma}_{j} \right)
   \; \rho \left( r_j \right) \, .
\end{eqnarray}
$G_F$ is the Fermi coupling constant,
$A$ is the number of nucleons,
$\gamma_j$ is the Dirac matrix, and
$C_{T}$ is a dimensionless coupling constant of the TPT interaction.
$C_{T}$ is equal to zero within the standard model,
but it is finite in some theories beyond the standard model
 of elementary particle physics.  
According to Dzuba~et~al~\cite{DzubaFlambaum:2009}
 \begin{eqnarray} 
\label{eq:C_T_ro}
   C_{T} < \hspace*{-.1cm}\bm{ {\sigma}}_{A} \hspace*{-.1cm}>  \; = \;
    \left<  C_{T}^{p} \sum_{p} {\bm{\sigma}_{p} }
    + C_{T}^{n} \sum_{n}  {\bm{\sigma}_{n}} \right> \, ,
\end{eqnarray}
where $\left< ... \right>$ represents averaging over the nuclear state
with the nuclear spin
$\bm{I}$. 
The nuclear charge density distribution
$\rho \left( r \right)$ is the 
normalized to unity two-component
Fermi function~\cite{grasp89}
\begin{equation}
\label{eq:distribut}
\rho(r)=\frac{\rho_0}{1+e^{(r-b)/a}}
\end{equation}
where $a$ and $b$ depend on the mass of the isotope.

The single-particle reduced matrix
elements $\left[ n_a \kappa_a \| \hat{t}^{k} \| n_b \kappa_b \right]$
in equation~(\ref{eq:mat_element_between_CSF})
 for the tensor - pseudotensor interaction
has the form

\begin{widetext}
\begin{eqnarray}
\label{eq:radial_integral_Tensor}
 \left[ n_a \kappa_a \|\hat{h}_{TPT}^{1} \| n_b \kappa_b \right] =
   \sqrt{2} \; G_{F} \; C_{T} \; < \hspace*{-.1cm}\bm{{\sigma}}_{A}
   \hspace*{-.1cm}> \;\left[ n_a \kappa_a \| 
   i \; \hat{\gamma}^{1} \; \rho \left( r \right) \| n_b \kappa_b \right] =
   \nonumber \\ [0.2cm]
   &  & \hspace*{-8.0cm}
 - \sqrt{2} \; G_{F} \; C_{T} \; < \hspace*{-.1cm}\bm{{\sigma}}_{A}
 \hspace*{-.1cm}> 
   \left\{ \left[ -\kappa_a \| \sigma^{1} \| \kappa_b \right] \;
   \int_{0}^{\infty}
   \hspace*{-.3cm} \ P_b Q_a  \; \rho\; dr \right. 
  \left. +\left[ \kappa_a \| \sigma^{1} \| -\kappa_b \right] \;
   \int_{0}^{\infty}
   \hspace*{-.3cm} \ P_a Q_b  \; \rho\; dr \right\} \, ,
\end{eqnarray}
where the single-particle angular reduced matrix elements can be expressed as:
\begin{eqnarray}
\label{eq:sigma1-ab}
   \left[ -\kappa_a \| \sigma^{1} \| \kappa_b \right]
    & = & 
  \frac{\left< l_b \frac{1}{2} 0 \frac{1}{2} | j_a \frac{1}{2} \right> 
  \left< l_b \frac{1}{2} 0 \frac{1}{2} | j_b \frac{1}{2} \right> -
  \left< l_b \frac{1}{2} 1-\hspace*{-.12cm}\frac{1}{2} | j_a \frac{1}{2}
  \right>
  \left< l_b \frac{1}{2} 1 -\hspace*{-.12cm}\frac{1}{2} | j_b \frac{1}{2}
  \right>}{\left< j_b 1 \frac{1}{2} 0 | j_a \frac{1}{2} \right>} ,
\end{eqnarray}

\begin{eqnarray}
\label{eq:sigma1a-b}
   \left[ \kappa_a \| \sigma^{1} \| -\kappa_b \right]
    & = & 
  \frac{\left< l_a \frac{1}{2} 0 \frac{1}{2} | j_a \frac{1}{2}\right>
  \left< l_a \frac{1}{2} 0 \frac{1}{2} | j_b \frac{1}{2}\right> -
  \left< l_a \frac{1}{2} 1 -\hspace*{-.12cm}\frac{1}{2} | j_a
  \frac{1}{2}\right>
  \left< l_a \frac{1}{2} 1-\hspace*{-.12cm}\frac{1}{2} | j_b
  \frac{1}{2}\right>}{\left< j_b 1 \frac{1}{2} 0 | j_a \frac{1}{2}\right>} .
\end{eqnarray}
\end{widetext}

%
\subsection{Pseudoscalar-scalar interaction}
\label{sectionSPS}

The interaction Hamiltonian for the
pseudoscalar-scalar ($PSS$) interaction between the electrons 
and the nucleus reads
\begin{eqnarray}
  \hat{ H}_{PSS} \; = \;\frac{-G_F\;C_{P}}{2\sqrt{2}m_{p}c} \;
                 {\sum_{j=1}^{N}} \;
  \gamma _{0} \; (\bm\nabla_{j}\rho \left( r_j \right) <
  \hspace*{-.1cm}\bm{ {\sigma}}_{A} \hspace*{-.1cm}>).
\end{eqnarray}
$C_{P}$ is dimensionless coupling constant of  the $PSS$ interaction.
Analogously to the $TPT$ interaction, $C_{P}$ constant is zero within
the standard model.
According to Dzuba~et~al~\cite{DzubaFlambaum:2009}
 \begin{eqnarray} 
\label{eq:C_P_ro}
   C_{P} < \hspace*{-.1cm}\bm{ {\sigma}}_{A} \hspace*{-.1cm}>  \; = \;
    \left<  C_{P}^{p} \sum_{p} \bm{ {\sigma}}_{p} 
    + C_{P}^{n} \sum_{n} \bm{ {\sigma}}_{n} \right> \, .
\end{eqnarray}

The single-particle reduced matrix
element $\left[ n_a \kappa_a \| \hat{t}^{k} \| n_b \kappa_b \right]$
in the equation (\ref{eq:mat_element_between_CSF}) for
the pseudoscalar-scalar interaction
has the form
\begin{widetext}
\begin{eqnarray}
\label{eq:radial_integral_SPS}
\left[ n_a \kappa_a \|\hat{h}_{PSS}^{1} \|         n_b \kappa_b \right] =
  -\frac{G_F\;C_{P}}{2\sqrt{2}m_{p}c}  < \hspace*{-.1cm}\bm{{\sigma}}_{A}
  \hspace*{-.1cm}>  
 \left[ n_a \kappa_a \| \gamma _{0} \;\nabla^{1}\rho \left( r \right) \| n_b
 \kappa_b \right] =
 \nonumber \\ [0.2cm]
   &  & \hspace*{-7.5cm}
 -\frac{G_F\;C_{P}}{2\sqrt{2}m_{p}c} < \hspace*{-.1cm}\bm{{\sigma}}_{A}
 \hspace*{-.1cm}>    
 \left[ \kappa_a \| C^{1} \|  \kappa_b \right]
 \int_{0}^{\infty}
 \hspace*{-.3cm} \left( P_a P_b \; - \; Q_a Q_b \right) \;\frac{d\rho}{dr} \;
 dr.
 \end{eqnarray}
\end{widetext}

%
\subsection{Schiff moment}
\label{sectionSchiff}

The Hamiltonian of this interaction ($H_{SM}$) can be expressed  as:
\begin{eqnarray}
\label{eq:H_SM}
   \hat{H}_{SM}  \; = \;
   \frac{3}{B} \;
   \sum_{j=1} ^{N} \; \left( \bm{ S}  \cdot \bm{r}_{j} \right)
   \; \rho \left( r_j \right) \, .
\end{eqnarray}

%
The Schiff moment $\bm{S}$ is directed along the 
nuclear spin $\bm{I}$ and $\bm{S} \, \equiv \, S \bm{I} / I$,
with $S$ being the coupling 
constant, and $B=\int_{0}^{\infty}\rho(r)r^{4}dr$.

The single-particle reduced matrix element
$\left[ n_a \kappa_a \| \hat{t}^{k} \| n_b \kappa_b \right]$ 
in expansion (\ref{eq:mat_element_between_CSF}) for SM 
can be factorized into reduced angular matrix element and radial integral
\begin{eqnarray}
\label{eq:radial_integral_Schiff}
 \left[ n_a \kappa_a \|\hat{h}_{SM}^{1} \|         n_b \kappa_b \right] =
           \frac{3}{B}
   \; S \; \left[ n_a \kappa_a \| 
   \hat{r}^{1} \; \rho \left( r \right) \| n_b \kappa_b \right] =
     \nonumber \\ [1ex]
   &  &  \hspace*{-8.0cm}
    \frac{3}{B}
   \; S \; \left[ \kappa_a \| C^{1} \| \kappa_b \right] \;
   \int_{0}^{\infty}
 \hspace*{-.3cm} \left( P_a P_b \; + \; Q_a Q_b \right) \; \rho \; r \; dr \, .
\end{eqnarray}

%
\subsection{Electron electric dipole moment}
\label{sectioneEDM}

The operator for the electron EDM interaction with magnetic field
of a nucleus can be expressed as:
\begin{eqnarray}
\label{op:HB}
  \hat{ H}_{B} \; = \;-id_{e}\;
                 {\sum_{j=1}^{N}} \;
  (\bm\gamma _{j} \; \bm B),
\end{eqnarray}
where $d_e$ represents the electron electric dipole moment,
and $\bm B$ the magnetic field of the nucleus.

The single-particle reduced matrix element
$\left[ n_a \kappa_a \| \hat{t}^{k} \| n_b \kappa_b \right]$ 
in expansion (\ref{eq:mat_element_between_CSF}) for
operator of electron EDM interaction with magnetic field of a nucleus
can be factorized into reduces angular matrix element and radial integral
\begin{widetext}
\begin{eqnarray}
\label{eq:radial_integral_eedm}
   \left[ n_a \kappa_a  
  \| h^{el}_{B} \| n_b \kappa_b \right]
    & = & 
   \nonumber \\ [0.2cm]
   &  & \hspace*{-2cm}
   \frac{d_{e}\mu}{2m_{p}c}\left\{ -3\left[ -\kappa_a \| C^{1}\| -\kappa_b
   \right] \;
   \int_{R}^{\infty}
    \hspace*{-.2cm} \frac{Q_a P_b}{r^3} \;  dr \right.
  \; - \; 3\left[ \kappa_a \| C^{1}\| \kappa_b \right] \int_{R}^{\infty} 
  \hspace*{-.2cm} \frac{P_a  Q_b}{r^3} \;  dr
%
  \; -\; \left[ -\kappa_a \| \sigma^{1}\| \kappa_b \right] \;
   \int_{R}^{\infty}
    \hspace*{-.2cm} \frac{Q_a P_b}{r^3} \;  dr
 \nonumber \\ [0.2cm]
   &  & \hspace*{-2cm}
   - \; \left[ \kappa_a \| \sigma^{1}\| -\kappa_b \right] \int_{R}^{\infty} 
   \hspace*{-.2cm} \frac{P_a  Q_b}{r^3} \;  dr
%
  \; + \; 2\left[ -\kappa_a \| \sigma^{1}\| \kappa_b \right] \;
   \int_{0}^{R}
   \hspace*{-.2cm} \frac{Q_a P_b}{R^3} \;  dr
  \left. \; + \; 2\left[ \kappa_a \| \sigma^{1}\| -\kappa_b \right]
  \int_{0}^{R} \hspace*{-.2cm} \frac{P_a  Q_b}{R^3}  \; dr \right\} ,
%
 \end{eqnarray}
\end{widetext}
where $R$ and $\mu$ represent the nuclear radius 
and nuclear magnetic moment, respectively.

We extended the GRASP2K~\cite{grasp2K:2013} package
for the calculation of the matrix elements 
(\ref{eq:mat_element_between_CSF}) 
and for the calculation of single-particle reduced matrix elements
(\ref{eq:radial_integral_EDM}),
(\ref{eq:radial_integral_Tensor}),
(\ref{eq:radial_integral_SPS}), 
(\ref{eq:radial_integral_Schiff}), and
(\ref{eq:radial_integral_eedm}).
The extension, presented in this work, includes subroutines for
calculation of matrix elements of
type $\left< i | \hat{H}_{int} |0 \right>$ from (\ref{eq:DAHint})
for
tensor-pseudotensor $\hat{H}_{TPT}$,
pseudoscalar-scalar $\hat{H}_{PSS}$, Schiff moment $\hat{H}_{SM}$, 
electron EDM interaction with nuclear magnetic field $\hat{H}_{B}$,
and electric dipole moment $\hat{D}_{z}$.

%
\section{MCDHF calculations}
\label{sectionMCDHFcalculations}

\subsection{MCDHF theory}
\label{sectionMCDHFtheory} 

We used the MCDHF approach to generate numerical representations
of atomic wave functions.
An atomic state function
(ASF)  $\Psi (\gamma P J M_{J})$
 is obtained as a linear combination of configuration state functions 
$\Phi (\gamma_{r} P J M_{J})$,
eigenfunctions of the parity $P$,
and total angular momentum operators $J^2$ and $M_{J}$:
%
\begin{equation}
\label{ASF}
\Psi (\gamma P J M_J) = \sum_{r} c_{r} \Phi (\gamma_{r} P J M_{J}),
\end{equation}
where $c_{r}$ are configuration  mixing coefficients.
The multiconfiguration energy functional was based on the 
Dirac-Coulomb Hamiltonian, given (in a.u.) by
\begin{equation}
\label{eg:MCDF}
\hat{H}_{DC}  =
\sum_{j=1}^N
\Big( c \boldsymbol \alpha_j \cdot \boldsymbol  p_j +
 (\beta_j-1)c^2 +V({r_j})\Big)
 + \sum_{j < k}^N
\frac{1}{r_{jk}},
\end{equation}
where $\boldsymbol \alpha$ and $\beta$ are the Dirac matrices, and
$p$ is the momentum operator.
The electrostatic electron-nucleus interaction,
$V({r_j})$,
has been generated from  a 2-parameter Fermi
nuclear charge distribution (\ref{eq:distribut}).
%
The effects of the Breit interaction, as well as QED effects, were
neglected, since they are expected to be small at the level
of accuracy attainable in the present calculations.

%
\subsection{Energy functionals}
\label{sectionEnergyFunctionals}

Several different methods of wave function generation were employed,
in order to test the dependence of the calculated atomic EDMs
on options available in the GRASP2K~\cite{grasp2K:2013} implementation
of the MCDHF method. 
One option is related to the variational energy functional
in the wave function optimisation procedure.
Two general forms of the energy functional are implemented
in the GRASP2K~\cite{grasp2K:2013} package:

\subsubsection{Extended Optimal Level}
\label{sectionEOL}
%
One-electron orbitals based on the Extended Optimal Level (EOL) form
are optimised to minimise the energy functional, which
is defined through the equation~(39) in reference~\cite{grasp89},
where generalised weights (equation~(40) in ref.~\cite{grasp89})
determine a specific atomic state ASF (or a set of ASFs).
Consequently, the orbitals in the EOL approach are optimal for a
specific atomic state ASF or a set of ASFs.

%
\subsubsection{Extended Average Level}
\label{sectionEAL}
One-electron orbitals based on the Extended Average Level (EAL) form
are optimised to minimise the
(optionally weighted) sum of energies of all ASFs which may be constructed 
from a given set of CSFs, so eventually it yields
an (optionally weighted) average energy of a set of atomic states.
This approach is computationally much cheaper, but usually less accurate than
the approach based on the EOL functional.

%
\subsection{Virtual orbital sets}
\label{sectionVirtualOrbitals}

%
The numerical wave~functions were obtained independently for the two
parities.
The calculations proceeded in two phases.
Spectroscopic (occupied) orbitals were obtained in the 
Dirac-Hartree-Fock approximation.
They were kept frozen in all subsequent calculations.
Then virtual (correlation) orbitals were generated
in several consecutive steps.
At each step the virtual set has been extended by one layer
of virtual orbitals.
A \textit{layer} is defined as a subset of virtual orbitals,
usually with different angular symmetries,
optimized simultaneously in one step,
and usually frozen in all subsequent steps.
In the present paper
three or four layers of virtual orbitals of each of the
 {\sl s, p, d, f, g} symmetries were generated.
At each stage only the outermost layer is optimized and the remaining
orbitals (spectroscopic as well as other virtual layers)
are kept frozen.
Virtual orbitals were generated in an approximation
in which all single and restricted double
substitutions from valence orbitals and a subset
of core orbitals
to subsequent layers of virtual orbitals
were included.
The restriction was applied to double substitutions in such a way that
only one electron was substituted from core
shells, the other one had
to be substituted from the valence shells
(i.e.~from {\sl 7s} shell in the case of even parity ground state
 of radium atom;
 {\sl 7s} and {\sl 7p} shells in the case of odd parity excited states
 of radium;
 {\sl 6s} and {\sl 6p} in the cases of mercury and ytterbium).
Four layers of virtual orbitals
were generated for each of the three elements -- Ra, Hg, Yb.
The combined contribution of the $n=3$ shells to
the hyperfine constants of the $7s7p$~$^1P$ state
was evaluated in a previous paper~\cite{Bieron:RaQ:2005}
and found to be negligible,
while the combined contribution of the $n=4$ shells 
was below 1 percent level.
Therefore in the present calculations the innermost core orbitals
$1s$, $2s$, $2p$, $3s$, $3p$, $3d$ of the radium atom
were kept closed for electron substitutions.
All other core orbitals, as well as valence orbitals, 
were subject to electron substitutions.
By similar argument,
the innermost core orbitals $1s$, $2s$, $2p$ of Hg and Yb 
were kept closed for electron substitutions.
The reader is referred to the
papers~\cite{Bieron:RaQ:2005,Bieron:Au:2009}
for further details of wave function generation.

%
\subsection{Non-orthogonal orbital sets}
\label{sectionNonOrtho}

%
The matrix elements of all interactions were calculated between the
 ground state $ns^2$ ($J=0$)  
and excited states  with total angular momentum $J=1$
and opposite parity
for $^{225}$Ra, $^{199}$Hg, and $^{171}$Yb.
In principle, the optimal wave functions for calculations
of EDM matrix elements
are obtained in the
Extended Optimal Level form 
(see section~\ref{sectionEOL} above) 
separately for each parity. 
%
%
%
The wave functions optimised separately for the ground and excited states
are built from independent sets of
one-electron orbitals.
The two sets are mutually non-orthogonal
and they automatically account for relaxation effects
involved in calculations of matrix elements between
different atomic states~\cite{Bieron:Ra3d2:2004,Bieron:RaSchiff:2009}.
%
On the other hand, the transition energies obtained
from
wave functions based separately optimised orbital sets may be less accurate
than  transition energies obtained from calculations based
on a common set of mutually orthogonal one-electron
orbitals.
The above situation often arises
when multiconfiguration expansions are tailored specifically
to include only those electron correlation effects that are important
for the one-electron expectation values.
For one-electron matrix elements involved in the present calculations
the dominant contributions arise from
single and restricted double substitutions.
We have not included the unrestricted double substitutions
i.e.~the electron correlation effects
 with dominant contributions to the total energy,
as well as higher order substitutions,
since their impact on EDMs is
indirect and usually small~\cite{Roberts:2013}.

We evaluated the effect of the relaxation of the wave functions by
performing two parallel sets of calculations based on a common
orbital set (orthogonal) and
on two separately optimised orbital sets (non-orthogonal), respectively.
Table~\ref{ortho-vs-nonortho} lists the atomic EDM for  $^{225}$Ra,
calculated in several approximations.
The first line (denoted 0(DF) in the first column)
lists the results obtained with
uncorrelated Dirac-Fock wave functions.
The following lines provide the results obtained with 
different numbers (1-4) of virtual orbital layers
included in the Virtual Orbital Set (VOS).
The number of virtual orbital layers in a given VOS is quoted in the
first column.
We skipped the 'orthogonal' calculation with four virtual orbital layers,
since the preceding lines show clearly that the effects of non-orthogonality
(i.e.~the relaxation of wave functions) are of the order of a
few percent, up to 11\% for the
interaction of the
electron electric dipole moment with the nuclear magnetic field
(eEDM entry in Table~\ref{ortho-vs-nonortho}).

The calculation of matrix elements in the non-orthogonal case
requires a transformation of one-electron orbitals from which the
wave functions of ground and excited states are built.
The program BIOTRA2~\cite{grasp2K:2013} was applied to
transform both wave functions to a biorthonormal
form~\cite{Malmqvist1986,Olsen1995} 
which then permits to use standard Racah algebra in
evaluation of matrix elements.

\begin{table}[htbp]
\begin{footnotesize}
\caption{Contributions to the atomic EDM from TPT, PSS, SM,
 and electron EDM interactions,
 calculated for $^{225}$Ra,
 using orthogonal (Orth) and non-orthogonal (Non-O) orbital sets.
The number VOS in the first column
is the number of virtual orbital layers.
Transition energies are experimental.  } 
\label{ortho-vs-nonortho}  
\begin{tabular}{llllllllllllllllll}
\hline\noalign{\smallskip}
\hline\noalign{\smallskip}
 &\multicolumn{2}{c}{TPT} &\multicolumn{2}{c}{PSS}
 &\multicolumn{2}{c}{SM}&\multicolumn{2}{c}{eEDM}  \\
\cline{2-3} \cline{4-5} \cline{6-7} \cline{8-9} \\
VOS&Orth & Non-O&Orth & Non-O&Orth &Non-O&Orth & Non-O\\
\colrule
\phantom{1}0(DF)&  -16.3   & -15.81  &-59.7  & -57.87  & -6.53  & -6.32  &-55.6  & -46.67 \\
\phantom{1}1    &  -14.5   & -15.51  &-53.3  & -57.09  & -6.28  & -7.01  &-48.1  & -43.69 \\
\phantom{1}2    &  -18.8   & -19.90  &-69.0  & -72.95  & -7.79  & -8.16  &-63.5  & -58.07 \\
\phantom{1}3    &  -19.9   & -20.68  &-70.3  & -75.83  & -8.27  & -8.59  &-66.9  & -60.13 \\
\phantom{1}4    &          & -20.28  &       & -74.42  &        & -8.63  &       & -58.45 \\
\colrule 
\end{tabular}
\end{footnotesize}
\end{table}

\subsection{Extended Optimal Level calculations}
\label{sectionExtendedOptimalLevels}

The final values of atomic EDMs, presented in the
Tables~\ref{tableRaHgYbTPTresults}, 
\ref{tableSPS},
\ref{tableSchiff}, and
\ref{tableeEDM_M},
were obtained with the Extended Optimal Level
optimisation procedure described in section~\ref{sectionEOL} above.
At each stage of generation of virtual orbital sets, a decision had
to be made with respect to the number of atomic levels included
in the variational energy functional.
Table~\ref{EOL_TPT_Ra_Ex}  presents the contributions
$d_{at}^{TPT}$
to the atomic EDM of $^{225}$Ra from the 
tensor-pseudotensor interaction 
(\ref{eq:H_T}).
The contributions 
from particular atomic states are listed in subsequent lines.
The radial wave functions were
optimised within the EOL procedure, with different numbers of EOL levels:
4, 6, 8, 10, or 12 levels,
as indicated in the first line of the Table~\ref{EOL_TPT_Ra_Ex}.
These data 
were obtained with experimental transition energies
quoted from the the NIST
Atomic Spectra Database
(NIST~ASD)~\cite{NIST-ASD}.

An inspection of the Table~\ref{EOL_TPT_Ra_Ex}
(the last line, denoted 'Sum All')
indicates that
the $d_{at}^{TPT}$ expectation value
becomes stable when eight or more levels are included in the 
Extended Optimal Level energy functional.
Analogous decisions were made for all virtual orbital sets,
as well as for the other two elements. 
The final calculations were made with varying numbers
of EOL levels, between 2 levels for uncorrelated Dirac-Fock
wave functions, with 6-8 levels in most correlated calculations,
and up to 13 levels in one case.

\subsection{Orbital contributions}
\label{sectionOrbitalContributions}


Another interesting conclusion arises from the analysis of contributions
of particular one-electron orbitals generated in the EOL optimisation
procedure.
The analysis presented in the Table~\ref{EOL_TPT_Ra_Ex}
was made with only one virtual orbital layer,
because the Extended Optimal Level
optimisation procedure described in section~\ref{sectionEOL} above
becomes unstable with the increasing numbers of virtual layers
and of EOL levels.
However, already at this level of approximation
the dominant contributions come from the singlet $7s7p$~$^1P$ and triplet 
$7s7p$~$^3P$ excited states. 
The states  $7s8p$~$^1P$ and $7s8p$~$^3P$, involving $8p$ orbital,
contribute 9\% and 3\%, respectively
(and their contributions partially cancel due to different signs).
All other states contribute less than one percent each.
The following lines present contributions of singlet and triplet
states generated by single or double electron substitutions from
the reference configuration $7s7p$ to the lowest available orbitals
$8s$, $8p$, and $6d$.
The line denoted 'Sum s-p' shows the contributions of the four dominant
states generated by single electron substitutions from the reference
configuration.
The line denoted 'Sum s-d' shows the sum of entries from the preceding two
lines of the $6d7p$ configuration;
the line 's-p+s-d' shows the sum of all preceding contributions.
The next six lines present the contributions of higher lying levels,
and the line 'Sum~D' show the sum of the contributions from these six
preceding lines.
The last line 'Sum~All' shows the total sum of all contributions
of all states listed in the preceding lines.
We present the partial sums ('s-p', 's-d', 's-p+s-d', and 'Sum~D')
to show their dependence on the number of EOL levels.
The contributions of individual levels are not very stable,
and in particular the small contributions may vary significantly,
but the partial sums are more stable, and the total sum ('Sum~All')
is strongly stabilized by the contributions from the dominant states.

It is interesting to make a comparison of Table~\ref{EOL_TPT_Ra_Ex}
with Table~VI from the reference~\cite{Latha:2013}.
In reference~\cite{Latha:2013}
the contributions from
$7s_{1/2}$-$7p_{1/2}$
and
$7s_{1/2}$-$8p_{1/2}$
single-particle matrix elements
(pairings in their language)
are of comparable sizes, -324.468 and -306.133, respectively,
while in our calculations
the relative sizes of the contributions from
$7s_{1/2}$-$8p_{1/2}$,
with respect to the contribution from
$7s_{1/2}$-$7p_{1/2}$
pairing,
are 9\% and 3\% for singlet and triplet states, respectively.
%
%
Also, there are differences with respect to the contributions
of higher symmetry orbitals. For instance, 
the contribution from
$d_{5/2}$ orbitals
is of the order of 4\%
(see TABLE~VII in reference~\cite{Latha:2013}),
while in our calculations 
the contributions from
$d_{5/2}$ orbitals
are below 1\%.

It is difficult to explain these differences, but one possible
explanation is due to differences in optimisation procedures
and radial shapes of one-electron orbitals which resulted
from these procedures, as discussed in the
section~\ref{sectionExtendedOptimalLevels}.
Different compositions of particular atomic states are
likely consequences of differences in radial bases.
The authors of the reference~\cite{Latha:2013}
used Gaussian basis sets,
while in our calculations we use numerical orbitals defined on a grid.
We do not have insight into the details of the calculations
presented in the reference~\cite{Latha:2013},
but their Gaussians are likely to be evenly distributed
over the entire configurational space.

Different theories use different methods of construction for atomic states.
A consequence of these differences 
is the fact, that comparisons of contributions from particular
atomic states or from individual one-electron orbitals  are
not  meaningful.
All excited and virtual orbitals generated in our calculations were
optimised with multiconfiguration expansions designed
for valence and core-valence electron correlation effects,
resulting in virtual orbital shapes with maximal overlaps with
valence and outer core spectroscopic orbitals.
Consequently, the correlation corrections to the wave function
are likely to be larger for the lower states included in the
Extended Optimal Level procedure.
%
We performed comparison calculations with virtual orbitals
generated with three different methods:
the Extended Average Level procedure,
as described in the 
section~\ref{sectionEAL};
with virtual orbitals
generated within the screened hydrogenic approximation;
and virtual orbitals from Thomas-Fermi potential.
As described in the
section~\ref{sectionEAL},
one-electron virtual orbitals generated with the EAL functional
are optimised to minimise the sum of energies of all states.
Hydrogenic and Thomas-Fermi virtual orbitals are not
variationally optimized, they just form orthogonal bases.
Our comparison calculations indicate, that calculations
based on
Extended Average Level, hydrogenic, and Thomas-Fermi virtual orbitals
converge slower than Extended Optimal Level calculations,
and the contributions of higher lying levels are larger,
compared to EOL results.

\begin{table}[htbp]
\begin{footnotesize}
\caption{$d^{TPT}_{at}$contribution to atomic EDM,
calculated with the EOL  method for 1st VOS,
using different numbers of optimized levels 
and experimental transition energies, in units 
 $\left(10^{-20} C_T \left< {\bf \sigma}_{A} \right> \left|e\right|
 \mbox{cm} \right)$,
for $^{225}$Ra. Numbers in brackets represent powers of 10.} 
\label{EOL_TPT_Ra_Ex}  
\begin{tabular}{lrrrrrrrrr}
\hline\noalign{\smallskip}
\hline\noalign{\smallskip}
 Levels &\multicolumn{1}{c}{4} &\multicolumn{1}{c}{6} &\multicolumn{1}{c}{8}
 &\multicolumn{1}{c}{10} &\multicolumn{1}{c}{12}  \\
\hline
\phantom{1}$7s$$7p$ $^3P$       &-5.00     &-4.46       &-4.63     &-4.59     &-4.63     \\
\phantom{1}$7s$$7p$ $^1P$       &-1.03[1]  &-8.80       &-8.70     &-8.69     &-8.57     \\
\phantom{1}$7s$$8p$ $^3P$       &          &0.39        &0.30      &0.33      &0.44      \\
\phantom{1}$7s$$8p$ $^1P$       &          &-1.12       &-0.96     &-1.01     &-1.24     \\
\hline                                                                                  
\phantom{1}Sum s-p              &-1.53[1]  &-1.40[1]    &-1.40[1]  &-1.40[1]  &-1.40[1]  \\
\hline                                                                                   
\phantom{1}$6d$$7p$ $^3D$       &2.53[-3]  &-7.72[-4]   &2.96[-2]  &-9.30[-2] &-6.91[-2] \\
\phantom{1}$6d$$7p$ $^3P$       &1.98[-1]  &-3.08[-2]   &-1.13[-1] &3.55[-2]  &7.34[-2]  \\
\hline                                                                                   
\phantom{1}Sum s-d              &2.00[-1]  &-3.16[-2]   &-8.33[-2] &-5.75[-2] &4.25[-3]  \\
\hline                                                                                   
\phantom{1}s-p+s-d              &-1.51[1]  &-1.40[1]    &-1.41[1]  &-1.41[1]  &-1.40[1]  \\
\hline                                                                                  
\phantom{1}$6d$$8p$ $^3D$       &          &            &-4.79[-2] &-9.44[-3] &-3.63[-3] \\
\phantom{1}$6d$$8p$ $^3P$       &          &            &-1.15[-1] &-4.90[-2] &-8.36[-2] \\
\phantom{1}$7p$$8s$ $^3P$       &          &            &          &-1.96[-2] &-2.20[-2] \\
\phantom{1}$7p$$8s$ $^1P$       &          &            &          &-6.02[-3] &-6.31[-3] \\
\phantom{1}$6d$$7p$ $^1P$       &          &            &          &          &-5.12[-3] \\
\phantom{1}$8s$$8p$ $^3P$       &          &            &          &          &1.50[-3]  \\
\hline                                                                                     
\phantom{1}Sum D                &          &            &-1.63[-1] &-8.41[-2] &-1.19[-1] \\
\hline                                                                                     
\phantom{1}Sum All              &-1.51[1]  &-1.40[1]    &-1.42[1]  &-1.41[1]  &-1.41[1]  \\
\hline                                                                                     
%
\end{tabular}
\end{footnotesize}
\end{table}

\subsection{Transition energies}
\label{sectionTransitionEnergies}

The summation in equation~(\ref{eq:DAHint})
runs over all excited states of appropriate parity and symmetry.
The contributions of higher lying levels are gradually decreasing,
since they are suppressed both by the energy denominators, as well as by
decreasing overlaps of one-electron radial orbitals,
entering integrals in the equations: (\ref{eq:radial_integral_Tensor}), 
(\ref{eq:radial_integral_SPS}), (\ref{eq:radial_integral_Schiff}), 
and (\ref{eq:radial_integral_eedm}). 
In numerical calculations they have to be cut off at certain
level of accuracy.
Except where indicated otherwise,
the results presented in the present paper
were computed with experimental transition energies
in the denominators of the matrix elements in equation~(\ref{eq:DAHint}).
The transition energies were calculated
%
from the NIST~ASD database~\cite{NIST-ASD} and
we include levels up to $6d7p~^{3}P$ 
for $^{225}$Ra, $6s8p~^{1}P$ for $^{171}$Yb, and $6s9p~^{1}P$ for $^{199}$Hg. 
However, several levels are missing in~\cite{NIST-ASD},
so we employed an approach, where those transition energies which
were not available, were replaced by the energies
calculated with one of the three different methods:
(1) using theoretical energies obtained from MCDHF approach;
(2) with the energy of the upper level replaced by the
    energy of the lowest excited state;
(3) with the energy of the upper level replaced by the
    experimental ionisation limit.
The choice was made between the above three options in 
case of each missing level, based on availability of a
reliable theoretical energy,
or alternatively on the proximity of the lowest excited state
or the experimental ionisation limit.
To verify this approach we performed test calculations,
where all three choices were used together.
Table~\ref{tableRaTPTenergies} presents the contributions
from the tensor-pseudotensor interaction to the atomic EDM
of radium isotope $^{225}$Ra.
Transition energies in Table~\ref{tableRaTPTenergies} were taken from:
MCDHF~RSCF calculation (RSCF),
MCDHF~RCI calculation (RCI),
experimental data (Expt),
experimental ionisation limit (ExIL),
experimental energy of the lowest excited level (Exp1).
The MCDHF~RSCF case was a self-consistent-field 
Extended Optimal Level calculation,
with 2, 6, 8 and 6 EOL levels
for DF, 1, 2, 3 and 4 VOS, respectively. 
The MCDHF~RCI case was a configuration-interaction calculation
with 100 levels included. Their differences indicate the
deviation incurred when the number of EOL levels is varied. 
It should be noted that experimental values of the energies
of the $7s7p$ levels were
used in all cases in columns 'Expt', 'ExIL', and 'Exp1'.
The lowest $nsnp$ levels yield the largest contributions to all EDM matrix
elements in the present calculations,
and their energies are available for all elements in question,
therefore replacements were made only for higher lying levels.
The number VOS in the first column of Table~\ref{tableRaTPTenergies} 
represents the number of virtual orbital layers.
These data indicate the sizes of errors, which may arise from replacing
experimental transition energies with 
experimental ionisation limit (ExIL) or
experimental energy of the lowest excited level (Exp1).
As can be seen, the deviation is less than 10\% in case of radium.
The deviations of the data obtained with calculated transition
energies are larger, due to the nature of the wavefunctions built
from non-orthogonal orbital sets, as explained in the
section~\ref{sectionNonOrtho} above.

\begin{table}[htbp]
\begin{footnotesize}
\caption{Tensor-pseudotensor interaction contributions to EDM, 
for $^{225}$Ra, 
in units
 $\left(10^{-20} C_T \left< {\bf \sigma}_{A} \right> \left|e\right|
  \mbox{cm} \right)$,
calculated with the EOL method and
compared with data from other methods.
Transition energies taken from:
MCDHF-RSCF calculation (RSCF),
experimental data (Expt),
MCDHF-RCI calculation (RCI),
experimental ionisation limit (ExIL),
experimental value of lowest excited level (Exp1).
(see text for explanation).
The number VOS in the first column
is the number of virtual orbital layers.}
\label{tableRaTPTenergies}  
\begin{tabular}{lrrrrr}
\hline\noalign{\smallskip}
\hline\noalign{\smallskip}
&\multicolumn{5}{c}{$^{225}$Ra}  \\
\cline{2-6} 
             VOS &   RSCF &    RCI &    Expt &   ExIL & Exp1  \\
\colrule
\phantom{1}0(DF) & -18.31 & -18.31 & -15.81  & -15.81 & -15.81 \\
\phantom{1}1     & -10.37 & -11.81 & -15.51  & -14.70 & -13.92 \\
\phantom{1}2     & -12.04 & -12.58 & -19.90  & -20.08 & -20.45 \\
\phantom{1}3     &        &        & -20.68  & -21.22 & -22.52 \\
\phantom{1}4     &        &        & -20.28  & -21.16 & -22.32 \\
\colrule 
Ref.~\cite{DzubaFlambaum:2009}(DHF)  & & & & & -3.5  \\
Ref.~\cite{DzubaFlambaum:2009}(CI+MBPT)  & & & & & -17.6  \\
Ref.~\cite{DzubaFlambaum:2009}(RPA)  & & & & & -16.7   \\
Ref.~\cite{Latha:2013}(CPHF)         & & & & &-16.585 \\
\colrule 
\end{tabular}
\end{footnotesize}
\end{table}

\subsection{Uncertainty estimates}
\label{sectionError}

%
Estimates of uncertainty in {\sl ab~initio} calculations are far more
difficult than the calculations themselves,
particularly in situations, where an atomic property is evaluated,
which has not been calculated before within the same approach for any
other element.
We can indicate possible sources of uncertainties, but their sizes
are difficult to estimate.
The possible sources of uncertainties are the following.
\begin{table}[htbp]
\begin{scriptsize}
\caption{Tensor-pseudotensor interaction contributions to EDM, 
calculated with the EOL method in different virtual sets,
in units
 $\left(10^{-20} C_T \left< {\bf \sigma}_{A} \right> \left|e\right|
  \mbox{cm} \right)$,
for $^{225}$Ra, $^{199}$Hg, and $^{171}$Yb,
compared with data from other methods.} 
\label{tableRaHgYbTPTresults}  
\begin{tabular}{lllllllllllllllllll}
\hline\noalign{\smallskip}
\hline\noalign{\smallskip}
&\multicolumn{3}{c}{$^{225}$Ra} &\multicolumn{1}{c}{$^{199}$Hg}
&\multicolumn{1}{c}{$^{171}$Yb}  \\
\cline{2-5} \cline{6-6} \cline{7-7} 
VOS                                       & Ex    & ExJL   &Ex1      &  Ex   & Ex \\
\hline
\phantom{1}0(DF)                          & -15.81 & -15.81 & -15.81 &  -6.15 & -3.31 \\ 
\phantom{1}1                              & -15.51 & -14.70 & -13.92 &  -4.86 & -1.94 \\ 
\phantom{1}2                              & -19.90 & -20.08 & -20.45 &  -5.70 & -3.71 \\ 
\phantom{1}3                              & -20.68 & -21.22 & -22.52 &  -6.10 & -4.03 \\ 
\phantom{1}4                              & -20.28 & -21.16 & -22.32 &  -5.53 & -4.24 \\ 
\hline                                                                                            
Ref.~\cite{DzubaFlambaum:2009}(DHF)       & -3.5  &        &         & -2.4   &  -0.70   \\
Ref.~\cite{Martensson:1985}(DHF)          &       &        &         & -2.0   &          \\
Ref.~\cite{DzubaFlambaum:2009}(CI+MBPT)   & -17.6 &        &         & -5.12  & -3.70    \\
Ref.~\cite{DzubaFlambaum:2009}(RPA)       & -17.6 &        &         & -5.89  & -3.37    \\
Ref.~\cite{Martensson:1985}(RPA)          &       &        &         & -6.0   &          \\
Ref.~\cite{Latha:2008}(RPA)               &       &        &         & -6.75  &          \\
Ref.~\cite{Latha:2013}(CPHF)              &-16.585&        &         &  -3.377&          \\
\hline 
\end{tabular}
\end{scriptsize}
\end{table}

\subsubsection{Electron correlation effects}
\label{sectionErrorEce}
%
%
%
In extensive, large-scale calculations the relative accuracy can
reach 1-5 percent, depending on the expectation value
in question (see eg.~\cite{Bieron:Au:2009,Bieron:RaQ:2005}).
An estimate of uncertainty associated
with the electron correlation effects can be obtained in several ways.
In the limit of very large number of virtual orbital layers 
an estimate of uncertainty may be related to oscillations of the
calculated expectation value plotted as a function of the 
size of the multiconfiguration
expansion~\cite{Bieron:Au:2009}.
In the present paper an estimate of the uncertainty was based
on the differences between the data obtained with the
largest two multiconfiguration expansions,
represented by 3 and 4 layers of virtual orbitals
in Tables~\ref{tableRaHgYbTPTresults},
\ref{tableSPS},
\ref{tableSchiff}, and
\ref{tableeEDM_M}.
We abstained from extending the virtual sets beyond fourth layer,
because there are several other possible sources of uncertainty
in the present calculations.
An inspection of the Tables
indicates that the differences between the last two lines
range between
0.47\% for the Schiff moment of Ra,
and 15.77\% for the Schiff moment of Hg (Table~\ref{tableSchiff}).
We may assume the latter as an estimate of uncertainty associated
with the neglected electron correlation effects.
\begin{table}[htbp]
\begin{footnotesize}
\caption{Pseudoscalar-scalar  interaction contributions to EDM,
calculated with the EOL method in different virtual sets in units 
 $\left( 10^{-23} C_P \left< \sigma_{A} \right> \left| e \right|
 \mbox{cm} \right)$
for $^{225}$Ra, $^{199}$Hg, and $^{171}$Yb,
compared with data from other methods.} 
\label{tableSPS}  
\begin{tabular}{llllllllllllllll}
\hline\noalign{\smallskip}
\hline\noalign{\smallskip}
VOS                                      &&\multicolumn{1}{c}{$^{225}$Ra} &&\multicolumn{1}{c}{$^{199}$Hg} &&\multicolumn{1}{c}{$^{171}$Yb}\\                             
\colrule                                                                                  
\phantom{1}0(DF)                         && -57.87 && -21.49 && -10.84 \\                      
\phantom{1}1                             && -57.09 && -17.16 && -6.31  \\                      
\phantom{1}2                             && -72.95 && -19.94 && -12.20 \\                      
\phantom{1}3                             && -75.83 && -21.53 && -13.26 \\                      
\phantom{1}4                             && -74.42 && -19.45 && -13.94 \\                      
\colrule                                                                                          
Ref.~\cite{DzubaFlambaum:2009}(DHF)      && -13.0  && -8.7   && -2.4      \\
Ref.~\cite{DzubaFlambaum:2009}(CI+MBPT)  && -64.2  && -18.4  && -12.4     \\
Ref.~\cite{DzubaFlambaum:2009}(RPA)      && -61.0  && -20.7  && -10.9     \\
\colrule 
\end{tabular}
\end{footnotesize}
\end{table}

\subsubsection{Wave function relaxation}
\label{sectionErrorWfr}
%
%
As explained in the section~\ref{sectionNonOrtho}
the effects of wave function relaxation were partially
accounted for in the present calculations, by
using non-orthogonal orbital sets for the opposite parities.
An inspection of Table~\ref{ortho-vs-nonortho}
indicates that the uncertainty
which may arise from wave function relaxation effects
is of the order of 10\%, although this estimate is based on relaxing only the
ASF wave function of the ground state on one hand, and the ASF wave functions
of all excited states taken together, on the other hand.
A more general, albeit far more expensive approach would be to
generate separate
atomic state functions for the ground state, as well as for each excited state,
implying non-orthogonality between all ASFs of both parities.

\begin{table}[htbp]
%
\begin{footnotesize}
\caption{Schiff moment contributions to atomic EDM,
calculated with the EOL method in different virtual sets, in units 
 $\left\{ 10^{-17} [S/(\left| e \right| \mbox{fm}^3)] \left| e \right|
 \mbox{cm} \right\}$,
 for $^{225}$Ra, $^{199}$Hg, and $^{171}$Yb,
compared with data from other methods.} 
\label{tableSchiff}  
\begin{tabular}{llllllllllllllll}
\hline\noalign{\smallskip}
\hline\noalign{\smallskip}
VOS                                          &&\multicolumn{1}{c}{$^{225}$Ra} &&\multicolumn{1}{c}{$^{199}$Hg} &&\multicolumn{1}{c}{$^{171}$Yb}\\
\colrule
\phantom{1}0(DF)                             && -6.32 && -2.46  && -1.54 \\                       
\phantom{1}1                                 && -7.01 && -2.45  && -0.88 \\                       
\phantom{1}2                                 && -8.16 && -2.23  && -1.83 \\                       
\phantom{1}3                                 && -8.59 && -2.98  && -2.05 \\                       
\phantom{1}4                                 && -8.63 && -2.51  && -2.15 \\                       
\colrule                                                                                          
Ref.~\cite{DzubaFlambaum:2009}(DHF)          && -1.8  && -1.2   && -0.42 \\
Ref.~\cite{DzubaFlambaum:2009}(CI+MBPT)      && -8.84 && -2.63  && -2.12 \\
Ref.~\cite{DzubaFlambaum:2009}(RPA)          && -8.27 && -2.99  && -1.95 \\
Ref.~\cite{Dzuba:2002}(RPA)                  && -8.5  && -2.8   &&       \\
Ref.~\cite{DzubaFlambaum:2007:76}(RPA)       &&       &&        && -1.9  \\
Ref.~\cite{Latha:2009}(CCSD)                 &&       && -5.07  &&       \\
\colrule 
\end{tabular}
\end{footnotesize}
\end{table}

\subsubsection{Energy denominators}
\label{sectionErrorEd}
%
%
%
As discussed in the section~\ref{sectionTransitionEnergies},
the summation in equation~(\ref{eq:DAHint})
runs over all excited states of appropriate parity and symmetry.
The NIST Atomic Spectra Database~\cite{NIST-ASD}
is of course finite, therefore
several levels with unknown energies had to be included in the
present calculations.
The uncertainty which may arise due to replacements described
in the section~\ref{sectionTransitionEnergies},
should not exceed 10\% in case of radium atom,
and we expect the same order of magnitude in case
of ytterbium and mercury.

\begin{table}[htbp]
\begin{footnotesize}
\caption{Contributions of electron EDM interaction with magnetic field
of nucleus,
to atomic EDM are calculated with the EOL method in different virtual sets,
in units ($d_e \times 10^{-4}$),
for $^{225}$Ra, $^{199}$Hg, and $^{171}$Yb,
compared with data from other methods.} 
\label{tableeEDM_M}  
\begin{tabular}{llllllllllllllll}
\hline\noalign{\smallskip}
\hline\noalign{\smallskip}
VOS                                     &&\multicolumn{1}{c}{$^{225}$Ra} &&\multicolumn{1}{c}{$^{199}$Hg} &&\multicolumn{1}{c}{$^{171}$Yb}\\
\colrule                                                          
\phantom{1}0(DF)                        && -46.67 && 13.41 && 5.37 \\              
\phantom{1}1                            && -43.69 && 9.58  && 3.17 \\              
\phantom{1}2                            && -58.07 && 12.22 && 5.72 \\              
\phantom{1}3                            && -60.13 && 12.80 && 6.09 \\              
\phantom{1}4                            && -58.45 && 11.45 && 6.44 \\              
\colrule                                                                                   
Ref.~\cite{DzubaFlambaum:2009}(DHF)     && -11    && 4.9   &&  1.0 \\
Ref.~\cite{Martensson:1987}(DHF)        &&        && 5.1   &&      \\
Ref.~\cite{DzubaFlambaum:2009}(CI+MBPT) && -55.7  && 10.7  && 5.45 \\
Ref.~\cite{DzubaFlambaum:2009}(RPA)     && -53.3  && 12.3  && 5.05 \\
Ref.~\cite{Martensson:1987}(RPA)        &&        && 13    &&      \\
\colrule
\end{tabular}
\end{footnotesize}
\end{table}

\subsubsection{Systematic errors}
\label{sectionErrorSe}
%
%
%
The possible sources of systematic errors include:
omission of double, triple, and higher order substitutions;
the effects of Breit interaction;
and QED effects.
The calculations of EDMs involve radial integrals of atomic
one-electron orbitals, but all these integrals
include factors in the integrands, which effectively cut off
the integrals outside the nucleus, so
the contribution to the integral comes from
within or in the vicinity of the nucleus.
Therefore an estimate of systematic errors can be made
by comparing the EDM calculations with hyperfine structure
calculations, where integrand in the form $r^{-2}$
appears in a one-electron integral, which in turn renders the
dominant contribution from the first half of the
radial orbital oscillation, i.e.~near the nucleus.
In certain cases in the hyperfine structure calculations 
the effects of double and triple substitutions can be quite
sizeable, of the order of 10-20\%, but they often partly cancel
and the net deviation is often smaller
than 10\%~\cite{Engels1993,Bieron:Au:2008}.
The effects of quadruple and higher order substitutions
are negligible.
The effects of Breit and QED are usually of the order
of 1-2 percent or less for neutral systems.

%
\subsubsection{Error budget}
\label{sectionErrorEb}
%
%
Based on the above estimates,
the relative root-mean-square deviation of the present calculations
yields 
$\sigma = 25$\%.

\section{Final results, discussion, and outlook}
\label{sectionFinal}

\subsection{Summary}
\label{sectionSummary}

Atomic EDMs arising from $(P,T)$-odd 
tensor-pseudotensor and pseudoscalar-scalar 
electron-nucleon interactions,
nuclear Schiff moment, and
interaction of electron electric dipole moment with nuclear magnetic field,
are presented in
Tables~\ref{tableRaHgYbTPTresults},
\ref{tableSPS},
\ref{tableSchiff}, and
\ref{tableeEDM_M},
for $^{225}$Ra, $^{199}$Hg, and  $^{171}$Yb.
The matrix elements and atomic EDMs were calculated 
using recently developed programs in the
framework of the GRASP2K code~\cite{grasp2K:2013}.
One of the objectives of the present calculations
was to test these programs.
Therefore the results are compared with the data
obtained by other methods:
random phase aproximations (RPA),
many-body perturbation theory and configuration interaction
technique (CI+MBPT),
coupled-cluster single-double (CCSD),
and  coupled-perturbed Hartree-Fock (CPHF) theory.
These methods are usually more accurate in calculations of properties
of closed-shell atoms.
%
%
An inspection of the Tables
indicates that the differences between our results and the data
obtained with the RPA 
methods~\cite{DzubaFlambaum:2009,Latha:2008,DzubaFlambaum:2007:76,%
Dzuba:2002,Martensson:1987,Martensson:1985}
range between
1.5\% for the Schiff moment of Ra (Table~\ref{tableSchiff}),
and 22.1\% for the tensor-pseudotensor of Hg
(Table~\ref{tableRaHgYbTPTresults}), 
all of them within the error bounds
estimated in the section~\ref{sectionErrorEb}
above.

Despite the reasonable agreement at the level of the correlated calculations,
very large differences should be noted at the uncorrelated levels,
DF (Dirac-Fock) in our calculations,
and DHF (Dirac-Hartree-Fock) in
references~\cite{DzubaFlambaum:2009} and~\cite{Martensson:1987}.
We used the different symbols to visually differentiate the
results obtained with different numerical codes, but 
the DF and DHF approximations are formally identical within
the Dirac-Fock theory, and they
should yield similar values, within numerical accuracies of
the Dirac-Fock codes.
A possible explanation of these large differences
may be the fact that in our (DF) calculations the
summation in equation~(\ref{eq:DAHint}) 
runs over only the two lowest excited states,
singlet $nsnp$~$^1P$ and triplet 
$nsnp$~$^3P$, which are generated at the Dirac-Fock level
of the 
GRASP2K code~\cite{grasp2K:2013}.
On the other hand,
in
references~\cite{DzubaFlambaum:2009} and~\cite{Martensson:1987}
the summation was probably carried over all excited states,
which can be constructed from a suitable set of virtual orbitals.
Otherwise we do not have an explanation.

%
%
Large differences at the level of the correlated calculations
should be noted between our results and the data
obtained with the CPHF theory~\cite{Latha:2013}.
The differences are:
18\% for TPT of Ra
and  39\% for TPT of Hg (see Table~\ref{tableRaHgYbTPTresults}).
%
%
The largest disagreement appears to be between the result
of the present calculations 
and the value obtained with the CCSD theory~\cite{Latha:2009} for
the Schiff moment of Hg (see Table~\ref{tableSchiff}).
The difference amounts to 102\%.
It is difficult to explain some of the abovementioned differences.
They may be due to different
orbital shapes,
orbital contributions,
and relaxation effects,
discussed in the sections~\ref{sectionOrbitalContributions}
and~\ref{sectionNonOrtho}, respectively.
%

%
Another objective of the present calculations was 
to test the methods of wave function generation,
as described in more detail in the section~\ref{sectionMCDHFtheory},
and of multiconfiguration expansions designed to account
for valence and core-valence electron correlation effects.
A reasonably good agreement of our results with
the data obtained within the RPA and CI+MBPT
methods~\cite{DzubaFlambaum:2009,Latha:2008,DzubaFlambaum:2007:76,%
Dzuba:2002,Martensson:1987,Martensson:1985}
seems to indicate that the multconfigurational model employed in
the present calculations accounts for the bulk of the
electron correlation effects.
With adequate computer resources,
these calculations may be extended in the future and include 
also core-core effects.
Based on the experiences with other atomic properties,
as well as on the present EDM calculations,
we expect that the accuracy of the EDM calculations may be
improved by a factor of ten, with respect to 
the current relative root-mean-square deviation of the order of
$25$\%.


\subsection{Outlook}
\label{sectionOutlook}

%
%
%
Several refinements are possible with respect to the methods
used in the present paper.
To account more accurately for the electron relaxation,
separate wave functions for the leading contributors to EDM may be generated.
A more general, albeit far more expensive approach
would be to generate separate
ASFs for the ground state, as well as for each excited state,
relaxing orthogonality of the orbital sets between all ASFs of both parities.

The expectation values
$d_{at}^{int}$ were calculated with theoretical (if reliable),
and  experimental (if available) transition energies,
as explained in the section~\ref{sectionTransitionEnergies}.
In fully correlated calculations theoretical
transition energies would have to be evaluated with all
single and unrestricted double substitutions. They would be
computationally much more expensive than those presented in
the present paper, but
possible with the currently available massively-parallel computers.
Electron correlation effects can also be accounted for using
the partitioned correlation function interaction
(PCFI) method~\cite{PCFI}, that allows 
contributions from single and unrestricted double substitutions
deep down in the atomic core to be summed up in a very efficient way.
In the near future
we will be able to perform fully {\sl ab initio} calculations
for atoms with arbitrary shell structures.
We are currently testing the latest version of the
GRASP package~\cite{grasp2K:2013}, with angular programs
providing full support for arbitrary numbers of electrons
in open {\sl spdf} shells. 


\begin{acknowledgments}

\noindent
The authors wish to thank the Visby program of
the Swedish Institute for a collaborative grant. 
%
\noindent
JB~acknowledges the support from
 the Polish Ministry of Science and Higher Education (MNiSW)
in the framework of the scientific grant No.~N~N202~014140
awarded for the years 2011-2014.
The large-scale calculations were carried out with the supercomputer Deszno
purchased thanks to the financial support of the European Regional
Development Fund in the framework of the Polish Innovation
Economy Operational Program (contract no. POIG.02.01.00-12-023/08).

\end{acknowledgments}

\bibliography{xet}

\end{document}